\documentclass[letterpaper, 10 pt, conference]{ieeeconf}
\IEEEoverridecommandlockouts
\usepackage{graphicx}
\usepackage{wrapfig}
 \usepackage{mathtools} 
\usepackage{amssymb}
\usepackage{amsmath,dsfont}
\usepackage{xcolor}
\usepackage{nicefrac}

\usepackage{float}

\newcommand{\extrawork}[1]{}

\usepackage{hyperref}
\begin{document}

\title{What should the encroaching supplier do in markets with some loyal customers? A Stackelberg Game Approach}
%
%
\author{Gurkirat Wadhwa \and
Veeraruna Kavitha}

\maketitle              
\begin{abstract}

Considering a supply chain (SC) with partial vertical integration, we attempt to seek answers to several questions related to the cooperation-competition based friction, abundant in such networks. 
Such an SC can represent a supplier with an in-house production unit that attempts to control an out-house  production unit via the said friction. The two production units can have different sets of loyal customer-bases and the aim of the manufacturer supplier-duo would be to get the best out of the two customer bases. 
  Our analysis shows that under certain market conditions, an optimal strategy might be to  allow both units to earn positive profits—particularly when they hold similar market power and when customer loyalty is high. In cases of weaker customer loyalty, however, the optimal approach may involve pressurizing the out-house unit to operate at minimal profits. Even more intriguing is the scenario where the out-house unit has a greater market power and customer loyalty remains strong; here, it may be optimal for the in-house unit to operate at a loss just enough to dismantle the downstream monopoly.
\end{abstract}

\newcommand{\Canticipate}[1]{}
\newcommand{\ignore}[1]{}
\newcommand{\ENMremoved}[1]{}
\newcommand{\Details}[1]{}
\newcommand{\extra}[1]{}
\newcommand{\jlt}[1]{}
\newcommand{\TR}[2]{#2}

\renewcommand{\nu}{w}

\newcommand{\y}{{\bf y}}


    




\newcommand{\p}{{\tilde p}}

\newtheorem{cor}{Corollary}

\newcommand{\eop}{{\hfill~$\Box$}}
\newcommand{\tus}[1]{{\color{red}{#1}}}
\newcommand{\symcase}[1]{{\color{blue}{#1}}}

\renewcommand{\a}{{\bf a}}
\newcommand{\C}{{\mathbb C}}

  \renewcommand{\H}{\mathbb{H}} 
  
\newcommand{\U}{{\mathcal U}}
\newcommand{\N}{{\mathcal N}}
\newcommand{\D}{{\mathcal D}}
\newcommand{\B}{{\mathcal B}}

\renewcommand{\S}{{\mathcal S}}
\newcommand{\E}{{\mathcal E}}

\newcommand{\A}{{\mathbb A}}
\renewcommand{\P}{{\mathbb P}}
\newcommand{\M}{{\mathbb M}}
\newcommand{\G}{{\mathbb G}}
\newcommand{\V}{{\mathbb V}}
\renewcommand{\S}{{\mathbb S}}
\newcommand{\Sj}{{{\mathbb S}_{j }}}
\newcommand{\So}{{{\mathbb S}_1}}
\newcommand{\St}{{{\mathbb S}_2}}
\newcommand{\Aset}{{\cal A}}
\renewcommand{\L}{{\cal L}}
\newcommand{\Sec}{{\cal S}}

\newcommand{\Mo}{{{\mathbb M}_1}}
\newcommand{\Mt}{{{\mathbb M}_2}}
\newcommand{\Mi}{{{\mathbb M}_i}}
\newcommand{\Mminusi}{{{\mathbb M}_{-i}}}
\newcommand{\Mj}{{{\mathbb M}_j}}
\newcommand{\Vi}{{{\mathbb V}_i}}

\newcommand{\Vot}{{{\mathbb V}_{12}}}
\newcommand{\Vto}{{{\mathbb V}_{21}}}

\newcommand{\Dbar}{\bar{D}}
\newcommand{\dbar}{\bar{d}}
\renewcommand{\b}{{\bf b}}
\newcommand{\indc}[1]{\mathds{1}_{\left\{ #1 \right \}} }
\renewcommand{\B}{{\mathcal B}}
\newcommand{\BR}{{\mathbb B}}
\newcommand{\lcm}{{\underline c}_m}
\newcommand{\les}{{\underline e}_s}
\newcommand{\LES}{{\underline E}_s}
\newcommand{\LCM}{{\underline C}_m}
\newcommand{\lcs}{{\underline C}_m} 

\newcommand {\lnu}{{\underline \nu}}
\newcommand{\lalpha}{{\underline \alpha}}
\newcommand{\lcf}{{{\underline C}_F}}

\newcommand{\fC}{f_\C}
\newcommand{\zero}{0_n}
\newcommand{\F}{{\cal F}}

\newcommand{\bp}{{\bf p}}
\newcommand{\bq}{{\bf q}}
\newcommand{\bbf}{{\bf f}}

\newcommand{\sG}{{\mbox{\fontsize{5.2}{5.2}\selectfont{$\G$}}}}
\newcommand{\sC}{{\mbox{\fontsize{5.2}{5.2}\selectfont{$\C$}}}}
\newcommand{\sM}{{\mbox{\fontsize{4.7}{5}\selectfont{$\M$}}}}
\newcommand{\sMi}{{\mbox{\fontsize{4.7}{5}\selectfont{$\Mi$}}}}
\newcommand{\sMj}{{\mbox{\fontsize{4.7}{5}\selectfont{$\Mj$}}}}
\newcommand{\sMminusi}{{\mbox{\fontsize{4.7}{5}\selectfont{${\mathbb M}_{-i}$}}}}
\newcommand{\sS}{{\mbox{\fontsize{5.2}{5.2}\selectfont{$\S$}}}}
\newcommand{\sSj}{{\mbox{\fontsize{5.2}{5.2}\selectfont{$\S_{j}$}}}}
\newcommand{\sSo}{{\mbox{\fontsize{5.2}{5.2}\selectfont{$\So$}}}}
\newcommand{\sSt}{{\mbox{\fontsize{5.2}{5.2}\selectfont{$\St$}}}}
\newcommand{\sMo}{{\mbox{\fontsize{5.2}{5.2}\selectfont{$\Mo$}}}}
\newcommand{\sMt}{{\mbox{\fontsize{5.2}{5.2}\selectfont{$\Mt$}}}}
\newcommand{\sL}{{\mbox{\fontsize{5.2}{5.2}\selectfont{$\L$}}}}
\newcommand{\sV}{{\mbox{\fontsize{5.2}{5.2}\selectfont{$\V$}}}}
\newcommand{\sH}{{\mbox{\fontsize{5.2}{5.2}\selectfont{$\H$}}}}
\newcommand{\sA}{{\mbox{\fontsize{5.2}{5.2}\selectfont{$\A$}}}}

\renewcommand{\C}{{\mathbb C}}
\newcommand{\x}{{\bf x}}
\newcommand{\bst}{{I}}
\newcommand{\I}{{\mathbb I}}

\newcommand{\GC}{{\P_{\sG}}}
\newcommand{\CHC}{{\P_{\sH}}}
\newcommand{\VC}{{\P_{\sV}}}
\newcommand{\ALC}{{\P_{\sA}}}

\newcommand{\DP}{{\bar D}}

 \newcommand{\pCHC}{\CHC}

 \newcommand{\pGC}{\GC}

 \newcommand{\pVC}{\VC}

\newcommand{\pALC}{\ALC} 

\newcommand{\RB}[1]{}






\newcommand{\pmax}{p_{mx}}
\newcommand{\qmax}{q_{max}}

\newtheorem{lemma}{Lemma}
\newtheorem{theorem}{Theorem}

\vspace{-2mm}

\section{Introduction}
\label{sec_intro}

Supplier encroachment refers to a strategic move where suppliers bypass traditional distribution channels, set-up an in-house production unit and   sell some finished-products directly to the end consumers, while continuing   supplying    to   lower echelon (downstream) agents.  This increasing trend of supplier encroachment (see e.g., \cite{arya2007bright,ha2022supplier,yoon2016supplier}), allows suppliers to exert greater influence over the downstream market and thereby gain more control over the supply chain (SC).
This trend is evident across various industries, transforming conventional SC dynamics. In the electronics industry,   companies like  Intel\href{https://www.intel.com/content/www/us/en/company-overview/company-overview.html/}, which traditionally supply electronic components, now have expanded their product offerings to provide end-to-end solutions.  In the automotive industry, major suppliers such as \href{https://www.boschautoservice.com/} Bosch and \href{https://www.continental-aftermarket.com/us-en} Continental now market parts and services directly to consumers, thereby enhancing their brand visibility and fostering closer customer relationships. 
Acer Inc., initially a supplier for IBM and Apple, leveraged this strategy to become one of the largest computer manufacturers worldwide by 2007 (\cite{nystedt2007acer}).
%
%

The trend is closely linked to another aspect namely  `vertical integration'  studied in SC literature (e.g., \cite{williamson1971vertical,simchi1999designing}). Vertical integration typically implies the integration of various units (across various echelons) into a single unit that   controls    multiple stages of production and distribution   (e.g., \cite{simchi1999designing,wadhwapartition,zheng2021willingness}).  The idea in most of this literature is to illustrate  the advantages of a centralized SC formed by complete integration of all  the manufacturers and the supplier.  Recently in \cite{wadhwapartition}, we showed that  for an SC supplying   essential products, a partial integration of one supplier and manufacturer is more stable (a unit that is not easily opposed by other collaborative arrangements) than the  centralized SC.  

The common feature in both the aspects mentioned above,  is a single unit that has capacity  spanning across multiple echelons. 
In 
 this study, we investigate one such  SC with one supplier and two manufacturers, where the supplier collaborates with one of the manufacturers resulting in a partial vertical integration, while competing with the other.  This study allows us to explore the optimal operating strategies for the vertical collaborating unit and there by derive it's worth,  when it acts as a leader by setting the wholesale price for the   materials supplied (to the out-house manufacturer) and by quoting another price to the end customers. This aspect is useful to study `stability' of  collaborating units.
 The current   paper derives the worth  under  fairly general conditions  compared to that in \cite{wadhwapartition}, which     focuses only on essential products.

An alternate interpretation of our SC is related to supplier encroachment. One can view the above arrangement as  a supplier  with an in-house production unit that also outsources   materials to an independent out-house production unit (or as a manufacturer with direct retail capability and another  retailer that  outsources production completely to the former). The primary goal in this context is to identify the market conditions under which it is advantageous for the supplier to operate  in both the roles.  
There are several other related   questions  that require attention. For example,    it might be  beneficial to shut-down  the in-house unit under certain conditions. Under certain other conditions, it might be  beneficial to operate its in-house unit at losses
 to inflate the demand of the products of the opponent, which in turn can become a substantially    more profitable venture.  Being a leader, it can also force the opponent to operate at negligible profits, if that becomes the optimal choice.  Our aim is also to investigate if such operating conditions can ever become optimal, and if so under what market conditions? The paper primarily
focuses on this aspect. The results of this paper can also be used to study coalition formation aspects under more general conditions than in \cite{wadhwapartition}.

\ignore{

Several other related questions emerge: for instance, under certain conditions, it may be optimal for the supplier to temporarily shut down its in-house unit. Alternatively, it may be beneficial for the supplier to operate its in-house unit at a loss if doing so increases the demand for the independent manufacturer's products, potentially leading to long-term profitability gains. Additionally, as a market leader, the supplier might impose conditions that force the independent manufacturer to operate at minimal profits if this aligns with optimal outcomes. This study aims to examine whether such operating conditions can be advantageous and, if so, under what specific market circumstances.
The paper primarily focuses on this aspect.

This problem can be seen as deriving the worth of a coalition formed by partial vertical cooperation between a supplier and one of the manufacturers, in a two manufacturer and one supplier two echoelon supply chain. 
{\color{blue}
The objective is to understand the optimal operating point for the vertical collaborating unit, when it acts as a leader by setting the wholesale price   for the raw-materials supplied to   the out-house production unit.}

{\color{blue} Alternatively one can view it as the scenario where  a supplier with an in-house production unit  is also   outsourcing it's raw material to an alternate production unit.   The primary objective in this context is to determine the market conditions  under which it is beneficial for the supplier to operate both in-house and out-house production units. 
There are several other interesting questions  that require attention. For example    it might be  beneficial to shut-down  the in-house unit under certain conditions. Under certain other conditions, it might be  beneficial to operate its in-house unit at losses
 to inflate the demand of the products of the opponent, which in turn can become a substantially    more profitable venture.  Being a leader, it can also force the opponent to operate at negligible profits, if that becomes the optimal choice.  Our aim is also to investigate if such operating conditions can ever become optimal, and if so under what conditions?
}
}

We consider a Stackelberg (SB) game framework, where the coalition of supplier and manufacturer  acts as the leader, and the out-house manufacturer is the follower.
Under some mild conditions on market potential and production, procurement and operation costs (which are essential for the survival of  the involved agents), we show the existence of  Stackelberg equilibrium.  By solving several sub-problems arising out of  various operating configurations, we derive  meaningful insights into this complex problem. 


The major findings of this study, some of which are supported by numerical illustrations are as follows: (i) when the two production units are of comparable 
strengths and  are not substitutable (where the customers are extremely loyal to their respective manufacturers), both of them derive strict positive profits at the optimal operating point of the supplier-manufacturer duo; (ii) more interestingly, at the optimal choice for the market with not-so loyal customers, the out-house manufacturer is compelled to operate at par (with almost zero profit margins);   and (iii) when the production units are of significantly different strengths, it is never optimal to allow both the production units to derive strict positive utilities; 
either it is optimal to operate the in-house   at losses, just sufficient to ensure the out-house is not a monopoly in the downstream market, or    to force the out-house unit to operate  with  negligible profit margins.

\subsubsection*{Literature Survey}

This kind of supplier-encroachment problem started with \cite{arya2007bright}, there are limited strands of literature thereafter, please refer to \cite{amirnequiee2024navigating,wang2013advantage} and the reference therein. 
In \cite{wang2013advantage}, the authors consider the original equipment manufacturer (OEM) making a  decision between an encroaching  supplier  (or  competitive manufacturer) and non-encroaching supplier(s). 
In \cite{amirnequiee2024navigating},
the authors again consider a similar variant, but with far more interesting features --- they also consider a two period game, that allows them to understand the future encroachment possibilities, and the future quality improvements of the suppliers. But to the best of our knowledge none of these models (or those considered in other SC based literature) consider a more realistic scenario with possible dedicated customer bases. The supplier and the manufacturer can have their individual reputations by virtue of which they can enjoy a loyal customer base, and such a provision in available in our model via the parameter $\varepsilon$. Further, we consider non-zero production costs as well as operational costs, which is again a more realistic aspect. As  a resultant of all these considerations, we have mathematical models (representing demands realized, and the utility functions, etc)  which are significantly more complex owing to  a number of discontinuities (our functions are at maximum piecewise concave/convex).

\ignore{
and  a more recent paper in this where the agents choose the sale quantity and then the price is decided according to the inverse demand function.

There is some limited literature studying different other forms of partial integration. 
Authors in \cite{wang2013advantage} examine a supply chain with an original equipment manufacturer (OEM) and a contract manufacturer (CM), where the OEM outsources part of its production to a competitive CM and the rest to non-competitive CMs. 
\ignore{The study explores different game-theoretic scenarios, such as the OEM acting as the Stackelberg leader in one scenario, and the CM as the leader in another, with wholesale price and outsourcing quantity as key variables.} 
In contrast, our study focuses on a supplier akin to a competitive CM, which has capabilities for both production and raw material supply, while the competing manufacturer fully depends on the supplier for raw materials. Unlike \cite{wang2013advantage}, we investigate scenarios in which it may be beneficial for the supplier to shut down its in-house production unit, thereby allowing only the independent manufacturer to operate in the market under certain conditions.
In \cite{kaya2011outsourcing}, the authors compare outsourcing with in-house production by examining contractual impacts on demand, focusing on the specific decision of outsourcing versus self-manufacturing. While their work concentrates on contract terms, this study focuses more on coalition value in a partially cooperative and competitive market setup, aiming to determine optimal operating conditions and pricing strategies for a supplier functioning as both a producer and a raw material supplier.

    A recent work in \cite{amirnequiee2024navigating} studies a two period model with a manufacturer and two suppliers, one with high process quality and other with low process quality . The manufacturer outsources it's production to the supplier and the supplier also makes it's self branded products. The supplier may decide to encroach in the downstream market which means that he will sell his self branded products in the downstream market. In contrast to this scenario, we consider the study of the supplier having an in-house production unit and also outsources raw material to an out- house manufacturer who is it's  rival in the downstream market. Our results differentiate in the way that they establish that the existence of future outsourcing
 opportunities strengthens the manufacturer’s leading role in the competition whereas we demonstrate that this supplier with an in-house production unit empowers it to make this manufacturer operate at negligible profits under certain market conditions.}

\ignore{
 The authors in \cite{wang2021supplier} study in which the supplier sells his products through a dual purpose retailer whose profit function consists of  his own profit and the consumer surplus where the second component can be assigned the weightage. The authors consider a inverse linear  demand  model and the supplier has an option to encroach but can also choose not to encroach if he finds it beneficial to do so while we also give the supplier this kind of flexibility in a sense that it may shut down it's in-house production unit if it finds it beneficial to do so. Our model again contrasts from this study as it considers the supplier selling his products through a dual purpose retailer while in our case , the role of the supplier is to outsource the raw materials to out-house manufcaturer while maintaining an in-house production unit.
 Their findings indicate that dual-purpose suppliers may reduce the need for deep SC collaboration with retailers, as direct consumer engagement allows for greater control. This shift could incentivize suppliers to bypass traditional collaborations while our study, however, suggests that collaboration through partial integration can create a stable SC structure, allowing the coalition to set optimal prices while maintaining competitive pressure. This stability supports the idea that partial integration may be more resilient than either full integration or complete independence in volatile markets.

To the best of our knowlege, no study considers the supplier }.

 \ignore{
In this study, we aim to determine:

The Stackelberg equilibrium for the supplier-manufacturer coalition,
Conditions under which both the coalition and the independent manufacturer should operate or where one may optimally exit the market,
The impact of these decisions on overall supply chain profitability.
By solving these sub-problems under various conditions, we seek to offer insights into strategic pricing decisions within supply chains involving partial vertical cooperation, competition, and complex market dynamics.

}

\ignore{
{
\subsubsection*{Literature Survey:}
 Authors in \cite{wang2013advantage} focus on quantity leadership in outsourcing, emphasizing output control to secure competitive advantages without direct consumer sales or pricing considerations while our work examines partial integration and its impact on supplier-led pricing and production strategies.


The authors in \cite{wang2013advantage}  investigates a supply chain comprising an original equipment manufacturer (OEM) and a contract manufacturer (CM),in which the 
 OEM outsources a part of its production to the competitive CM and the remainder to non-competitive CMs and then differnet types of games are studied such as a sequential game with the OEM as the
 Stackelberg leader, and a sequential game with the CM as the Stackelberg leader which depend upon the outsourcing quantity and
 wholesale price. In our work we consider a supplier which can be compared to competitive CM who is capable of both the production of the product as well as the raw material supply and the competitive manufacturer has to rely on the supplier completely for the raw material supply for the production as opposed to \cite{wang2013advantage}. Further, the authors prove that when the competitive CM sets the
 wholesale price, it always sets it sufficiently low to
 allow both parties to coexist in the market while in our work we prove that under some conditions there may be a possibility of the supplier to shut down it's own production unit and allowing only the competitive manufacturer to operate which may be beneficial for the supplier.

In \cite{kaya2011outsourcing}, the authors compare outsourcing and in-house production contracts based on effort-dependent demand. The key difference is that the authors focus on the contractual implications  of the outsourcing and in-house production decisions while we focus on the game-theoretic coalition analysis which helps us to study the strategic decisions of the supplier who is  capable of both producing and supplying the raw material.}}

 \vspace{-3mm}

\section{Problem statement}
\vspace{-1mm}
Consider  a partially integrated SC with one   supplier $S$  that collaborates with one of the manufacturers (say $M_i$)  by forming a coalition $\Vi = \{M_i, S\}$ and competes with another (say $M_j$). We assume the supplier and it's coalition to quote their prices  first: wholesale price $q$   to $M_j$ for raw materials and price $p$ for final product to the end-customers. Thus  we have a
  Stackelberg game with  coalition $\Vi$ as the  leader and the   manufacturer  $M_j$ as the  follower.  The 
 manufacturer  $M_j$ 
  \begin{wrapfigure}{r} {0.3\textwidth}
 \vspace{-4mm}
    \centering
\includegraphics[trim = {.3cm 0.8cm .7cm 0.1cm}, clip, scale = 0.2 ]{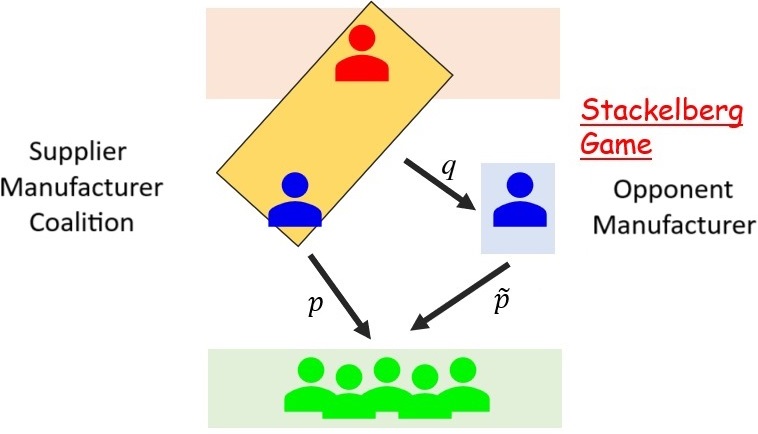}
\vspace{-2mm}
{\caption{Model Description} 
\label{fig:model}}
\vspace{-5mm}
\end{wrapfigure}
 can choose to operate by quoting a price $\tilde{p}$ to the end customers for the finished product using  the  raw material supplied by $\V_i$ (see Figure \ref{fig:model}). It can  also choose not to operate represented by action $n_o$, depending upon $q$ and the market response; with such a choice, the corresponding unit is completely shut-down and incurs zero utility (zero profit and  zero   cost).

 \noindent{\bf Market Response:}  The demand attracted by any manufacturer depends upon the price quoted for the finished product, for example, that attracted by
 manufacturer $M_{i}$ is given by  (see \cite{wadhwapartition,zheng2021willingness} for similar models):
 \vspace{-2mm}
\begin{eqnarray}\label{eqn_demand_vc_coal}
D_\sMi =  (\dbar_\sMi - \alpha_\sMi p +\varepsilon\alpha_\sMj \p)^{+},
\end{eqnarray}
where the different influencing factors are as  below:
\begin{itemize}
\item $\dbar_\sMi$  is the dedicated market potential of manufacturer~$M_i$,
\item $\alpha_\sMi p$ is the fraction of demand lost by $M_i$ due to
 its quoted   price $p$, sensitized by  parameter $\alpha_\sMi$ (here $\alpha_\sMi$ can be  a representative of the reputation of $M_i$),
 
 \item The demand is positive as long as the term inside
 $(.)^+$ is positive; else, the demand is zero.

 \item  \ignore{the last component $\varepsilon g_j(\p) $ is due an `unhappy' fraction of the loyal customer base of the out-house manufacturer $M_j$ and is explained in the immediate following.} $\varepsilon \alpha_\sMj
 \p$ is the fraction of customer base of $M_j$
 that rejected $M_j$ (due to its quoted price $\p$) and got converted as customers of $M_i$.
\end{itemize}

\ignore{
The demand $D_\sMj$ attracted by manufacturer $M_j$ 
has exactly similar structure with respective parameters, ($\dbar_\sMj, \alpha_\sMj $), the fraction of the loyal customers of $M_j$ that are unhappy with price $\p$  exactly equal $g_j(\p) = 
\varepsilon \alpha_\sMj \p$ and an $\varepsilon$  fraction among these unhappy customers seek service from $M_i$; additionally observe $g_j(\p)$ can at maximum be $\dbar_\sMj$, see  \eqref{eqn_demand_vc_coal}.} 

The parameter $\varepsilon$ represents the substitutability of the manufacturers. 
When $\varepsilon \approx 1$, the manufacturers are substitutable and the customers can buy the product from any of the manufacturers. On the other hand, when $\varepsilon \approx 0$, the manufacturers are not substitutable, i.e.,  the customers are loyal and choose  to buy the product only from `their'   manufacturers. 

\noindent {\bf Utilities:} 
We begin with the utility of out-house manufacturer $M_j$. When it does not operate, represented by indicator  $\F^c_\sMj = \indc{\p = n_{o}}$, it derives zero utility. 
When it operates (represented by $\F_\sMj$), it attracts demand as in \eqref{eqn_demand_vc_coal} and then the revenue derived equals the demand times the price minus the expenses. 
 Thus the utility of the manufacturer $M_j$  equals:
\begin{eqnarray}
 U_\sMj &=& \left(D_\sMj \left(\p - q - C_\sMj \right) {\cal F}_{\sV_i} - O_\sMj \right)\F_\sMj \\ \mbox{ with } D_\sMj &=&  (\dbar_\sMj - \alpha_\sMj \p +\varepsilon\alpha_\sMi p)^{+},  \label{Eqn_Umj}
\end{eqnarray}
where    $C_\sMj$ represents the  production cost per unit and $O_\sMj$ represents the operating cost. The profit of manufacturer
 $M_j$ is zero   when  the
 supplier does not operate (represented by indicator ${\cal F}_{\sV_i}^c$). 

The utility of $\V_i$ due to  demand $D_\sMi $  attracted by its manufacturer will have similar structure.   
Additionally, the demand  $D_\sMj$ attracted by $M_j$ also contributes towards the revenue of $\V_i$  (as it  supplies raw material).
  In all,  the utility of the coalition $\V_i$ is given by,

\vspace{-4mm}
{\small\begin{eqnarray}
 U_{{\sV}_i} &=& \big(D_\sMi\left(p-C_\sMi -C_\sS \right)\F_\sMi \nonumber \\
 &&\hspace{5mm}+ D_\sMj \left(q-C_\sS \right)\F_\sMj
 - O_\sS - O_\sMi \big)\F_{{\sV}_i} \label{Eqn_Util_V},
\end{eqnarray}}%
where  $C_\sS$ represents the raw material procurement cost  (per unit) and $C_\sMi$, $O_\sS$ and $O_\sMi$ have similar interpretations. The coalition $\V_i$ can choose to shut    in-house production (or it's   manufacturer $M_i$) if it deems   advantageous,   represented by action $(p, q)$ with $p=n_o$, and hence the inclusion of the  flag   $\F_\sMi := \indc{p \ne n_o}$ in  \eqref{Eqn_Util_V}; alternatively it might find it beneficial to not operate at all, indicated by  
$ \F^c_{{\sV}_i} = 1-\indc{q  \ne n_o}$.

We need to choose an upper bound for the prices. Observe that the demand attracted by any manufacturer (say $m$) gets zero, even after considering the maximum possible fold-back from the other manufacturer (say $-m$), if 
$
p_m >  \nicefrac{ (\dbar_m + \varepsilon \dbar_{-m})} {\alpha_m}. 
$
Thus we set $\pmax =\nicefrac{(\dbar_\sMi + \varepsilon\dbar_\sMj)}{\alpha_\sMi}$  and $\p_{mx} =\nicefrac{(\dbar_\sMj + \varepsilon\dbar_\sMi)}{\alpha_\sMj}$ as the maximum prices respectively  for $M_i$ and $M_j.$ 
\textit{We assume that if any agent is indifferent between the action~$a = n_o$ and an $a \ne n_o$,   the agent prefers operating choices.} This consideration is inspired from the practical scenarios (see \cite{wadhwapartition}). We further consider   the following assumptions as in \cite{wadhwapartition}, which ensures none of the  agents find it beneficial not to operate:
\begin{itemize}
    \item [{\bf A.1}] Assume the market potentials are sufficiently high, i.e., 

    \vspace{-4mm}
{\small\begin{eqnarray*}
    \dbar_\sMi  &\ge& \alpha_\sMi(C_\sS + C_\sMi) +2\sqrt{\alpha_\sMi(O_\sS + O_\sMi)}  \mbox{ and }\\
\dbar_\sMj  &\ge& \alpha_\sMj(C_\sS + C_\sMj)\\
&&+ 2\max\{\sqrt{2\alpha_\sMj (O_\sS + O_\sMi)},\sqrt{ \alpha_\sMj O_\sMj}\} .    
    \end{eqnarray*}}

  \item [{\bf A.2}] Assume, 
   $\varepsilon \le \left(\nicefrac{2\sqrt{\alpha_\sMj O_\sMj}}{\alpha_\sMj C_\sMi}\right).$ 
\end{itemize}
Assumption  {{\bf A.1}} ensures that the market potentials of both the manufacturers are sufficiently high compared to production, procurement and operating costs  (see \cite{wadhwapartition} for similar details). We will observe that    $\V_i$   finds it optimal to operate (i.e $\F^*_{{\sV}_i} =1$),  which is important for meaningful analysis.
 Assumption  {{\bf A.2}} is required for some technical reasons; besides, in general the operating costs are significantly large compared to (per-unit) production costs and hence the assumption would automatically be satisfied (note here $\varepsilon \le 1$).  
 
\subsection{Preliminary analysis and discussions}

\noindent {\bf   Best response of   $M_j$:}
We  begin  by obtaining the best response of the follower, the out-house manufacturer $M_j$,  when the Stackelberg leader (coalition $\Vi$)  declares $  (p, q)$. In particular we consider the case with ${\cal F}_{\sV_i} =1$, or when $\Vi$ decides to operate. This  response of $M_j$  is governed by the  following optimization problem (observe from   \eqref{Eqn_Umj} that $D_\sMj$ depends upon $(p,q)$):

\vspace{-4mm}
{\small\begin{eqnarray*}
U^{*}_\sMj(p,q) = \sup_{\p \in \{n_o, [0, \p_{mx}] \}} \Big( D_\sMj (\p - C_\sMj - q) - O_\sMj \Big) \indc{\p \ne n_o}.
\end{eqnarray*}}%
Such a problem is considered in \cite[Lemma 4]{wadhwapartition}.  By similar concavity arguments, the best response exists and equals:

\vspace{-4mm}
{\small\begin{eqnarray}
\p^*(p,q) = \min\left\{ \frac{\dbar_\sMj + \varepsilon \alpha_\sMi p}{2 \alpha_\sMj} + \frac{C_\sMj + q}{2}, \p_{mx} \right\} \indc{q \le \theta(p)} \nonumber \\ + n_o \indc{q > \theta(p)}, \mbox{ with, } \hspace{3mm} 
\label{Eqn_opt_policy_Mj} 
\end{eqnarray}
\begin{eqnarray}
 \theta(p) := \left\{ 
\begin{array}{lll}
\frac{\dbar_\sMj +\varepsilon\alpha_\sMi p -\alpha_\sMj C_\sMj - 2\sqrt{\alpha_\sMj O_\sMj}}{\alpha_\sMj} & \mbox{if } p < p_{sw}, \\
\frac{\dbar_\sMj + \varepsilon\dbar_\sMi - \alpha_\sMj C_\sMj}{\alpha_\sMj} - \frac{\alpha_\sMj O_\sMj}{\alpha_\sMj (\varepsilon\alpha_\sMi p - \varepsilon\dbar_\sMi)} & \mbox{else,}
\end{array}
\right.  \hspace{-85mm}   \label{Eqn_feasible_Regioin_Mj}\\ 
& p_{sw} := \frac{\dbar_\sMi}{\alpha_\sMi} + \frac{\sqrt{\alpha_\sMj O_\sMj}}{\varepsilon\alpha_\sMi}, \mbox{ and recall, } p_{mx} = \frac{\dbar_\sMi + \varepsilon \dbar_\sMj}{\alpha_\sMi}.  \hspace{2mm} \label{Eqn_psw}
\end{eqnarray}}
In the above $p_{sw}$ represents a switching point ---  if the price of in-house $M_i$ is above $p_{sw}$,
the optimal price of the opponent $M_j$ is clamped at the maximum possible value $\p_{mx}$.
Further,
  $M_j$ may not find it beneficial even to operate if the price $q$ quoted for raw materials is high (this happens when $q > \theta (p) $ in \eqref{Eqn_opt_policy_Mj}). Interestingly,  this  also depends upon the price $p$ quoted by the in-house manufacturer $M_i$   towards the end-product. More interestingly $M_j$ can tolerate a larger $q$ if the price $p$ is higher (observe $\theta(p)$ increases with $p$) --- a large part of loyal customer-base of in-house $M_i$ can improve market opportunities for $M_j$ 
\ignore{(observe $\varepsilon g_i (p)$ seek products from $M_j$, and that $p \mapsto g_i (p)$ is increasing).} (observe from  \eqref{Eqn_Umj} that $\varepsilon \alpha_\sMi p$  fraction of customers seek products from $M_j$, and it is increasing in $p$).

\noindent {\bf Choices of   coalition $\V_i$: }    The  coalition $\V_i$ comprising of in-house manufacturer $M_i$ and supplier $S$  has several advantages, as the vertical cooperation provides it  multiple choices.  
 
 \noindent$ \bullet${\bf [Eliminate downstream competition]} The existence of in-house manufacturer in $\V_i$ provides it an option to operate in monopolistic manner when it is possible to attract a large fraction of `unhappy' loyal customers of the opponent $M_j$; this is possible probably when the manufacturers are substitutable to a good extent, i.e., when   $\varepsilon$ is large.  In  this case, it can completely eliminate 
    $M_j$ (by quoting exorbitantly large $q$)   and operate in the monopolistic manner in the downstream market with the combined market potential,  $\dbar_\sMi + \epsilon \dbar_\sMj$. 
    
 \noindent$ \bullet${\bf [Shut down the in-house]} If either the market potential of the in-house manufacturer is low or when its reputation is not very good (when $\alpha_\sMi$ is more, its  customers are highly sensitive to price $p$), or when those factor of the out-house are significantly better, then $\V_i$ has an  option to completely shut its in-house production unit $M_i$.  Such a choice can reduce the competition for opponent $M_j$ which in turn can become beneficial for $\V_i$ --- it may have an option to sell large amount of raw material (as market $D_\sMj$ attracted by $M_j$ can be large)  at good/optimal prices. 

However it may not be beneficial to allow the out-house to operate in a monopolistic manner; like-wise it may not be beneficial to completely eliminate out-house $M_j$ unless the two production units are completely substitutable (in an ideal world with $\varepsilon=1$).  In such cases, there are other choices for $\Vi$ which we describe next and which are the focus of this paper.

 \noindent$ \bullet${\bf [Co-existence]}   
    In this scenario, both    $\Vi$   and out-house  $M_j$ operate; rather $\Vi$ allows both to operate.  By virtue of this, it can charge sufficiently large (optimal) price $q$ for raw materials, which (probably) leaves few choices for $M_j$ --- the latter then has to quote larger prices $\p$  to survive in the downstream market. This   facilitates  $\V_i$ to benefit from both the worlds, because of the `unhappy' loyal customers ($\varepsilon \alpha_\sMi p$) of $M_j$ that seek product from $M_i$ as well as from the  high profits derived by selling the raw material to $M_j$ at large~$q$. Basically it chooses optimal $( p, q)$ that provides the best combined utility as a Stakelberg leader, while competing with the out-house manufacturer $M_j$ in the downstream market. 
 There are several sub-possibilities for $\Vi$  under co-existence:
    \begin{itemize}
        \item {\bf [Operate both profitably]} The coalition $\V_i$ quotes the price pair $(p,q)$ such that both production units derive non-zero profits. 
        
         \item {\bf [In-house operates at losses]} Alternatively $\V_i$ can operate it's in-house production unit at losses (by quoting large $p$), if that could fetch it a   larger revenue by just supplying to $M_j$; basically it might be beneficial not to allow the out-house to operate in  monopolistic manner in the downstream  market, by expending towards operating its in-house.  
         \item {\bf [Out-house forced to operate at par]} In this case, the coalition$\V_i$ quotes the price $q$ to the manufacturer $M_j$ such that this manufacturer operates but gets \textit{zero revenue}-- this means that the coalition $\V_i$ quotes $q$ large but sufficient to keep the out-house manufacturer operate at par.
    \end{itemize}

   \ignore{ In this case, it would chose a $(p,q)$ such that $\alpha_\sMi p > \dbar_\sMi + \alpha_\sMj \p^*(p,q)  $ and hence such that $D_\sMi = 0$ and then optimizes the following:
    \begin{eqnarray*}
        \sup_{p, q, \ s.t., \ D_\sMi = 0, \ q \le \theta(p) }  \left(\dbar_\sMj + \varepsilon \alpha_\sMi p - \alpha_\sMj \p^* (p,q) \right)(q- C_\sS) - O_\sMi - O_\sS \\
                \sup_{p, q, \ s.t., \ D_\sMi = 0, \ q \le \theta(p) }  \left(\dbar_\sMj + \varepsilon \alpha_\sMi p - \alpha_\sMj \left (e_1' + \frac{q}{2} +   \frac{p\varepsilon \alpha_\sMi}{2 \alpha_\sMj} \right ) \right)(q- C_\sS) - O_\sMi - O_\sS \\
                                \sup_{p, q, \ s.t., D_\sMi = 0, \ q \le \theta(p) }  \left(\dbar_\sMj + \frac{ \varepsilon \alpha_\sMi p}{2} - \alpha_\sMj \left (e_1' + \frac{q}{2}     \right ) \right)(q- C_\sS) - O_\sMi - O_\sS
    \end{eqnarray*}
    Thus the optimizers are $p^* = p_{max}$, the maximum possible value and $q^*$ satisfies:
    $$
    q^* = 
  \frac{  \dbar_\sMj + \frac{ \varepsilon \alpha_\sMi p_{max}}{2} - \alpha_\sMj  e_1'  + \frac{\alpha_\sMj C_\sS}{2} } {\alpha_\sMj}
    $$
    Thus the optimal utility  in this regime is given by:
    \begin{eqnarray*}
        \frac{ \left (\dbar_\sMj + \frac{ \varepsilon \alpha_\sMi p_{max}}{2} - \alpha_\sMj  e_1'       - \frac{\alpha_\sMj}{2} C_\sS  \right )^2 }{2 \alpha_\sMj }
    \end{eqnarray*}
    }

 \vspace{-1mm}

The main aim of this work is to analyze the optimal choice of the coalition $\V_i$ among  various sub-regimes of coexistence. The comparison with the other two regimes namely elimination of downstream competition and  shut down the in-house requires separate attention.  
In \cite{wadhwapartition}, while deriving the worth of  the partial vertical cooperation partition under essentialness conditions ($\varepsilon \to 1$ and $\alpha \to 0$),  we already discovered  that co-existence is optimal as  compared  to elimination of DS competition or shutting down the in-house production. There is a possibility that the answers could be similar for other market conditions, but that would be considered as a part of future work.  
%
%
%
We now begin with the main theme of the paper, the analysis of the co-existence regime. 

\vspace{-3mm}

\section{ Co-existence }

It is a Stackelberg game under the co-existence scenario with $\Vi = \{S, M_i\}$ as  the leader and $M_j$ as the follower. For any given $(p,q)$, the joint-price policy of $\Vi$, the optimal utility of out-house manufacturer $M_j$  is given by:

\vspace{-4mm}
{\small
\begin{eqnarray}
U^*_\sMj(p,q) = \big( \left( \dbar\sMj - \alpha_\sMj \p^{*}  + \varepsilon \alpha_\sMi p \right)^{+} \left( \p^{*}  - C_\sMj - q \right)  \nonumber \\
- O_\sMj \big) \indc{\p^{*} \ne n_o},
\label{Eqn_Opt_for_Mj}  
\end{eqnarray}}%
\ignore{
\begin{eqnarray}
U^*_\sMj(p,q) = \big (  \left( \dbar_\sMj \hspace{-1mm}- \alpha_\sMj \p^{*} + \varepsilon\alpha_\sMi p \right)^{+} \hspace{-1mm}\left   (\p^{*}  - C_\sMj - q\right)\nonumber\\ &&- O_\sMj \bigg) \indc{\p^* \ne n_o} \hspace{3mm},   
\label{Eqn_Opt_for_Mj}
\end{eqnarray}}%
where $\p^{*}=\p^*(p,q)$, the optimizer of out-house manufacturer $M_j$, is given by \eqref{Eqn_opt_policy_Mj}.
We are interested in obtaining the optimal utility under co-existence, where the utility for any $(p,q)$, for which $\p^* \ne n_o$, is given by:

\vspace{-4mm}
{\small\begin{eqnarray}
    U_\sV (p, q) &=& \left ( \dbar_\sMi - \alpha_\sMi p  + \varepsilon \alpha_\sMj \p^*(p, q) \right )^+ (p - C_\sMi - C_\sS) \nonumber \\
    && \hspace{-13mm}+ \left ( \dbar_\sMj + \alpha_\sMi \varepsilon p  -  \alpha_\sMj \p^*(p, q) \right )^+ (q- C_\sS)  - O_\sMi - O_\sS . \hspace{5 mm} \label{eqn_util_co-exist_given_pq}
\end{eqnarray}}

Thus
the feasible region for co-existence (possible only when   $M_j$ is also operating, and we include the possibility of $\Vi$ operating at losses), using \eqref{Eqn_opt_policy_Mj}-\eqref{Eqn_feasible_Regioin_Mj}  is given by:
\vspace{2mm}
\begin{eqnarray*}
{\cal F}_{co} &:= & \{ (p, q) \in (0, \infty)^2 :   q \le \theta(p)   \}.
\end{eqnarray*}

Towards  optimizing \eqref{eqn_util_co-exist_given_pq} with respect to $(p,q) \in {\cal F}_{co}$, first
consider the following `unconstrained' optimization problem, which resembles \eqref{eqn_util_co-exist_given_pq} but for $(\cdot)^+$ operators, and   when   $\p^*(p,q)  <  \p_{mx}$:
\begin{equation}
\hspace{-10mm} 
\resizebox{0.6\textwidth}{!}{ 
 $\begin{aligned}
    & \sup_{p,q} U(p,q) \quad \mbox{where, }  \\
    \hspace{8mm}U(p, q) &= \left( \dbar_\sMi - \alpha_\sMi p + \varepsilon \alpha_\sMj \left( \frac{\dbar_\sMj + \varepsilon \alpha_\sMi p}{2 \alpha_\sMj} + \frac{C_\sMj + q}{2} \right) \right) (p - C_\sMi - C_\sS) \\
    &\hspace{-10mm}+ \left( \dbar_\sMj + \alpha_\sMi \varepsilon p - \alpha_\sMj \left( \frac{\dbar_\sMj + \varepsilon \alpha_\sMi p}{2 \alpha_\sMj} + \frac{C_\sMj + q}{2} \right) \right) (q - C_\sS) - O_\sMi - O_\sS.\label{eqn_util_co-exist_uc}
  \end{aligned}$
  }
\end{equation}

The proof for the existence of the  optimizer of this unconstrained optimization problem is given in Appendix A.
Let $(p^*_{co}, q^*_{co} )$ represent its optimizer (which are derived  in     in the proof of Theorem \ref{thm_Fco_positive} provided in Appendix A) and equals:

{\small\begin{eqnarray}
    p^{*}_{co} &=& \hspace{-2mm} -\frac{2w_3w_4 - w_2w_5}{4w_1w_3 - w_2^2}, \quad
    q^{}_{co} = -\frac{w_2 p^{}_{co} + w_5}{2w_3}, \quad \mbox{with,} \nonumber\\
    w_1 &=& \hspace{-2mm}-\alpha_\sMi \left(1-\frac{\varepsilon^2}{2} \right), 
    w_2 = \frac{\varepsilon\left(\alpha_\sMi + \alpha_\sMj\right)}{2}, 
    w_3 = -\frac{\alpha_\sMj}{2}, \nonumber\\
   w_4 &=&\hspace{-2mm} \frac{\scriptstyle 2\dbar_\sMi + \varepsilon\dbar_\sMj + \varepsilon\alpha_\sMj C_\sMj - \varepsilon\alpha_\sMi C_\sS + 2\alpha_\sMi\left(1-\frac{\varepsilon^2}{2}\right)(C_\sMi + C_\sS)}{2}, \nonumber\\
    w_5 &=& -\frac{\varepsilon\alpha_\sMj \left(C_\sMi + C_\sS\right)}{2} + \frac{\left(\dbar_\sMj - \alpha_\sMj C_\sMj + \alpha_\sMj C_\sS\right)}{2}. \ \hspace{4mm}
    \label{Eqn_ws}
\end{eqnarray}}
The above optimal pair $(p_{co}^*, q_{co}^*)$ can become the optimizer for the original co-existence objective function \eqref{eqn_util_co-exist_given_pq}. 
However,  \eqref{eqn_util_co-exist_given_pq} is different from the `unconstrained' function  \eqref{eqn_util_co-exist_uc} in some sub-regimes of the co-existence region ${\cal F}_{co}$. So in the quest towards the optimal co-existence policy, one also needs to find the optimizer(s) in the sub-regimes where the two differ.  In all, we will have partition of ${\cal F}_{co}$ into many sub-regimes, such  that the objective functions  \eqref{eqn_util_co-exist_given_pq} and   \eqref{eqn_util_co-exist_uc} match
\begin{wrapfigure}{r} {0.2\textwidth}
    \centering
    \vspace{-4mm}
\includegraphics[trim = {.2cm 0.3cm .7cm 0.3cm}, clip, scale = 0.1]{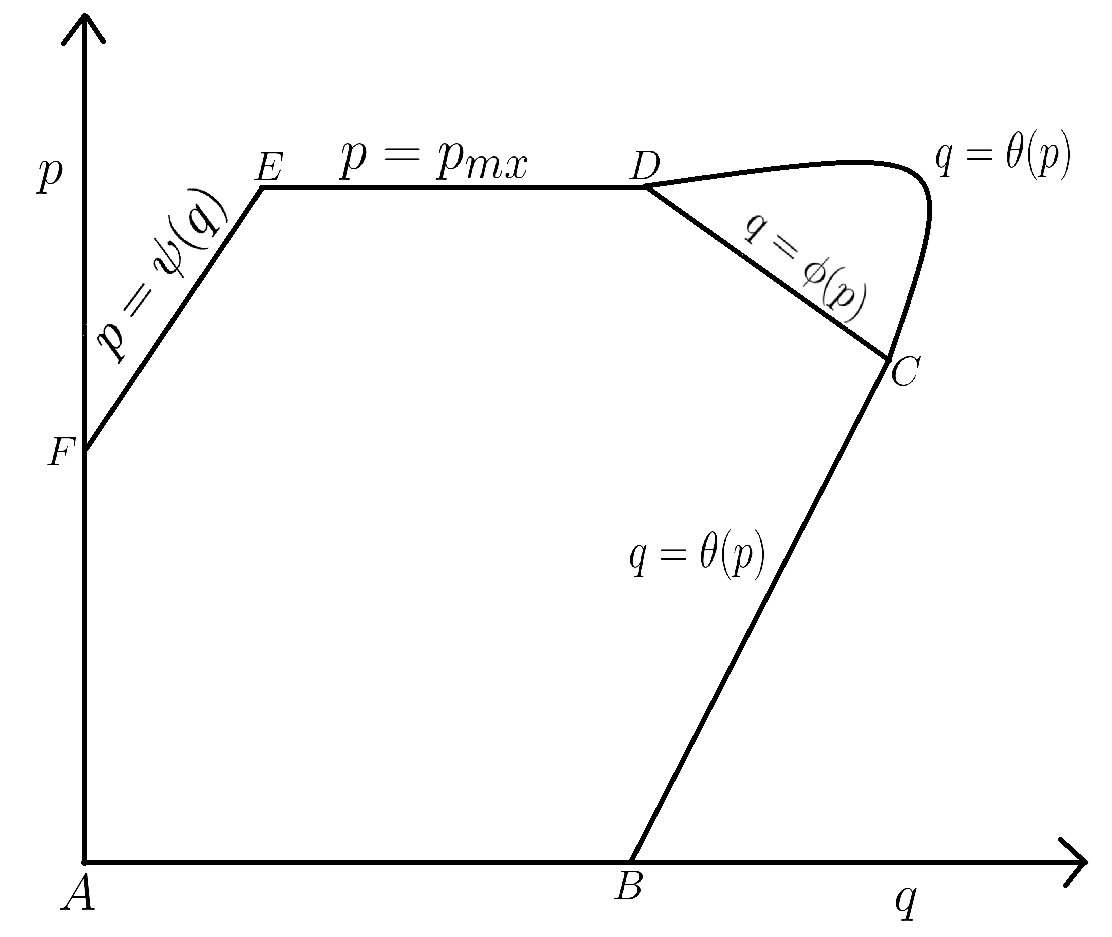}
{\caption{A representative Feasible Region, ${\cal F}^+_{co}$} 
\label{fig:feasible region}}
\vspace{-4mm}
\end{wrapfigure}
 in the first sub-regime, while they differ in the remaining   sub-regimes.   We now consider them one after the other.  Interestingly three of these sub-regimes align with our initial discussion on the choices of $\Vi$, however, an extra boundary line $\{p=\pmax\} \cap {\cal F}_{co}$  of ${\cal F}_{co}$ also becomes important and requires attention.


\vspace{2mm}
\subsection{ Operate Profitably for both}

Consider a sub-region \textit{in the interior of which both the manufacturers derive strictly positive utility}; such a choice also ensures that \eqref{eqn_util_co-exist_given_pq} 
equals \eqref{eqn_util_co-exist_uc}, as will be evident from below. We first identify this sub-regime,  the interior of which  should satisfy the following:
\begin{itemize}
    \item  includes the pair of prices $(p,q)$, for which the optimal price $\p^*(p,q)$ of the out-house manufacturer is less than $\p_{mx} = \nicefrac{(\dbar_\sMj + \varepsilon \dbar_\sMi)}{\alpha_\sMj}$; such a regime from \eqref{Eqn_opt_policy_Mj}-\eqref{Eqn_feasible_Regioin_Mj} is given by  $\{ (p,q) :  q < \phi (p) \}$ where $\phi (\cdot) $ is defined below 
  \begin{eqnarray}\label{Eqn_phi_st_p}
   \phi (p)  := \frac{\dbar_\sMj + 2\varepsilon\dbar_\sMi - \varepsilon\alpha_\sMi p - \alpha_\sMj C_\sMj}{\alpha_\sMj}; 
    \end{eqnarray}
    the boundary of  such a regime is the  straight line,  $\mathbb{L}_1 :=  \{ q =  \phi (p)\} $; 
this condition ensures $\p^*$ in \eqref{eqn_util_co-exist_given_pq}  matches with its counterpart in \eqref{eqn_util_co-exist_uc};

    \item includes the pair of prices $(p,q)$, for which  the $\Vi$ coalition derives strict positive utility from in-house production unit also;  this is the sub-region where   $\dbar_\sMi + \varepsilon\alpha_\sMj \p^{*}(p,q) - \alpha_\sMi p >  0$; such a region, further  within  $\{  q < \phi (p) \}$, is given by $\{    q < \phi (p) \mbox{ and } p < \psi (q) \}$, with $\psi (\cdot)$ defined below: 
  \begin{eqnarray}
  \label{Eqn_psi_q}
  \psi (q)  :=  \frac{ \left ( 2 \dbar_\sMi + \varepsilon  \dbar_\sMj + \varepsilon \alpha_\sMj  (C_\sMj+q ) \right )}{(2-\varepsilon^2)\alpha_\sMi} ;     
  \end{eqnarray}

  observe here that  the   straight  line,     $\mathbb{L}_2 := \{p = \psi (q)  \}$, bounds the region  of interest only when  
    it is also bounded by $\mathbb{L}_1$ and these constraints ensure $(\cdot)^+$ terms in \eqref{eqn_util_co-exist_given_pq} are positive and hence match with the counter parts in \eqref{eqn_util_co-exist_uc};

    \item  the pair of prices $(p,q)$  which ensure  co-existence  (the out-house also operates) belong to    $\{q \le \theta(p)\}$ with $\theta(\cdot)$ as in \eqref{Eqn_feasible_Regioin_Mj};  
within a region bounded by lines, $\mathbb{L}_1$ and $\mathbb{L}_2$,     
       this region can be  bounded by  straight line (the first row of  \eqref{Eqn_feasible_Regioin_Mj} is already  to the left of $\mathbb{L}_1$, so sufficient to  consider second row of~\eqref{Eqn_feasible_Regioin_Mj}),
       {\small{
    $$
    \mathbb{L}_3 =\hspace{-1mm}  \left \{\hspace{-1mm}q = \theta(p) =   
      \frac{\dbar_\sMj + \varepsilon\alpha_\sMi p -\alpha_\sMj C_\sMj - 2\sqrt{\alpha_\sMj O_\sMj}}{\alpha_\sMj} \hspace{-1mm}\right \};
    $$}}

    \item  and finally bounded by $\mathbb{L}_4$, which represents the horizontal line of maximum price, $\{ p = \pmax \}$. 
    
\end{itemize}
 Such a sub-region (actually its closure),  represented by ${\cal F}^+_{co}$,  is the region in the positive quadrant,  bounded by all the lines $\mathbb{L}_1, \mathbb{L}_2, \mathbb{L}_3$ and $\mathbb{L}_4$ (see polygon ABCDEF in  Figure \ref{fig:feasible region} for one representative scenario). To summarize:
\begin{eqnarray} 
    {\cal F}^+_{co} =  \bigg \{ (p, q) \in  [0,\infty)^2  :    p \le \min \{\pmax, \psi(q)  \} \mbox{ and } \nonumber \\
    q \le \min \left \{\theta (p), \phi (p) \right  \} \bigg \}.   \label{Eqn_Fco_plus}
\end{eqnarray}
It is immediate that we have the following:
\begin{eqnarray}
     U_{\sV, co}^* = \max \left \{\hspace{-2mm} \max_{(p,q) \in {\cal F}_{co}^+ } U_\sV (p,q),   \max_{(p,q) \in {\cal F}_{co} \setminus {\cal F}_{co}^+ } U_\sV (p,q)     \right \}.
\end{eqnarray}
We first analyze the first term under {{\bf A.1}} and {{\bf A.2}}. 
\begin{theorem}
\label{thm_Fco_positive}
Assume {{\bf A.1}}-{{\bf A.2}}. 
(i)  If $(p^*_{co}, q^*_{co})$ is   in the interior of ${\cal F}_{co}^+$ then, 
\begin{eqnarray}
\label{eqn_vc_opt_f_co_+}
\max_{(p,q) \in {\cal F}_{co}^+ }U_\sV(p,q) = U_\sV(p^*_{co}, q^*_{co}).
  \end{eqnarray} 
(ii) If  $(p^*_{co}, q^*_{co})$ is not in the interior of ${\cal F}_{co}^+$, the   optimal  utility across ${\cal F}_{co}^+$ is at one of the non-empty boundaries, excluding the $\{q=0\}$ and $\{p=0\}$ lines:
\begin{eqnarray}
\max_{(p,q) \in {\cal F}_{co}^+ } U_\sV (p,q) =  \max_{l \in \{1, 2, 3, 4\}}  \left \{ \max_{ (p,q) \in {\cal F}^+_{co} \cap \mathbb{L}_l    }   U_\sV (p,q) \right \}.  
  \end{eqnarray} 
In the above, by convention, the maximum of an empty set is set to zero. 
\end{theorem}

{\bf Proof:} is provided in Appendix A. \eop

Next we analyze the region
${\cal F}_{co} \setminus {\cal F}_{co}^+ $.
It comprises of several sub-regions which we elaborate next. 

\vspace{2mm}
 \subsection  {  In-house   operates at loss}
\label{sec_in-house_at_loss}

 This is the sub-region in which the coalition $\V_i$ allows its in-house $M_i$ to operate while incurring losses. Basically the coalition $\V_i$ quotes a very large price $p$
resulting in zero demand for itself --- this could be beneficial for $\V_i$,  as such a tactic could create large opportunities via  the market  captured by the out-house $M_j$. 
  We denote this region as ${\cal F}_{co}^{ls}$, which is   given  by:
    \begin{eqnarray*}
        {\cal F}^{ls}_{co} &=&  \left \{ (p,q) \in {\cal F}_{co} :  p \le \pmax, p > \psi(q) \right \} \nonumber \\
      &=&  \left \{ (p,q) \in {\cal F}_{co} :  p \le  \pmax, p > \psi(q), q \le  \theta(p) \right \} . 
    \end{eqnarray*}
From \eqref{Eqn_psi_q},  $\psi$ is an increasing function and so \textit{ this regime is non-empty only when $\psi(0) < \pmax$}.
    The co-existence utility \eqref{eqn_util_co-exist_given_pq} of $\V_i$, specially represented by $U_{ls}(p,q)$ for this sub-case, equals (see \eqref{Eqn_opt_policy_Mj}):
    \begin{eqnarray*}
     U_{ls} (p,q) &=&  \left(\dbar_\sMj - \alpha_\sMj \p^{*}(p,q)+ \varepsilon\alpha_\sMi p \right)\left(q- C_\sS \right) \\ &&- O_\sMi - O_\sS 
     \mbox{ for all } (p,q) \in {\cal F}_{co}^{ls}.
    \end{eqnarray*}
Let the optimal utility under this be represented by $U_{ls}^{*}$.  From \eqref{Eqn_opt_policy_Mj},  it is straight forward to show that the   optimizer of the function (which is strictly increasing in $p$ for any fixed $q$), is given by $(\pmax,q^{ls,*})$ where $q^{ls,*}$ is the solution of the following optimization problem (once again, as $\psi$ is an increasing function):
\begin{eqnarray*}
      U_{ls}^{*} &=& \hspace{-2mm}\max_{q :(\pmax,q)\in {\cal F}_{co}^{ls} } \hspace{-2mm}U_{ls}(\pmax, q) =        \max_{q \le  u_{ls} } U_{ls}(\pmax, q),\\ \mbox{ with } 
        u_{ls} &:=& \min \left \{  \max\{0, \psi^{-1} (\pmax)\} , \theta(\pmax) \right \},   \nonumber
    \end{eqnarray*}
    where  using \eqref{Eqn_phi_st_p} we have for any $q \le u_{ls}$ (by convention   $[a,b] = \emptyset$ when $a > b$),
 
\begin{eqnarray*}
  U_{ls}(\pmax, q) &=& - O_\sMi - O_\sS \\
  && \hspace{-15mm} + \Bigg( \frac{\dbar_\sMj (1 + \varepsilon^2) + \varepsilon \dbar_\sMi - \alpha_\sMj (q + C_\sMj)}{2} \indc{q \le \phi(\pmax)} \\
  && \quad + \varepsilon^2 \dbar_\sMj \indc{q \in [\phi(\pmax), u_{ls}]} \Bigg) (q - C_\sS).
\end{eqnarray*}

    One can solve the first term without including the effect of $\indc{ q \le \phi(\pmax) } $ term to obtain ${\tilde q}^*$ as the optimizer (given below)  and then   derive the overall  optimizer for this sub-case as below:
    {\small{
    $$
       \hspace{-2mm} q^{ls, *} =  {\tilde q}^*    \indc{  {\tilde q}^* \le \min\{u_{ls}, \phi(\pmax) \}} +   u_{ls}  \indc{  {\tilde q}^* >  \min\{u_{ls}, \phi(\pmax) \}}      
  $$
  }}
  where
  $$
  {\tilde q}^* =  \left(\frac{\dbar_\sMj(1+\varepsilon^2) + \varepsilon\dbar_\sMi - \alpha_\sMj C_\sMj}{2\alpha_\sMj} + \frac{C_\sS}{2}\right).
  $$
Thus the optimal utility in this sub-regime (when non-empty) is given by:
\begin{eqnarray}
    U_{ls}^* = U_{ls} (\pmax, q^{ls, *} )  \indc{ \psi(0) < \pmax}. \label{Eqn_opt_loss_util}
\end{eqnarray}

  \subsection{ Operate at maximum price }
  We  now consider the sub-regime inside  $\{p=\pmax\}$, a  boundary of ${\cal F}_{co}^+$ that can potentially house optimizer   on $\mathbb{L}_4$ line. This line can become a part of the boundary along the segment  $[ l_{mx}, r_{mx} ]   $,  only when  $l_{mx}  <  r_{mx}$,  where from definitions,
  \begin{eqnarray}
   l_{mx} &:=& \scalebox{0.95}{$\max\left \{ \psi^{-1} (\pmax), 0 \right \}$} \nonumber\\
   &=&  \scalebox{0.95}{$\max\left \{ \frac{ -\varepsilon \dbar_\sMi + (1-\varepsilon^2) \dbar_\sMj - \alpha_\sMj C_\sMj }{ \alpha_\sMj}, 0 \right \}$} \hspace{-20mm}   \label{Eqn_psi_inv_pmax}\\
 r_{mx} &:=& \scalebox{0.95}{$\bar{q}(\pmax)$} \nonumber\\
 &=& \scalebox{0.95}{$\left \{ 
    \begin{array}{lll}
       \frac{\dbar_\sMj (1+\varepsilon^2) + \varepsilon \dbar_\sMi -\alpha_\sMj C_\sMj - 2\sqrt{\alpha_\sMj O_\sMj}}{\alpha_\sMj}   \hspace{-2mm} & \mbox{ if } \pmax \le  p_{sw} \\[8pt]
       \frac{\dbar_\sMj (1-\varepsilon^2)+ \varepsilon\dbar_\sMi - \alpha_\sMj C_\sMj}{\alpha_\sMj} & \mbox{ else. }
    \end{array}
\right .$}   \label{Eqn_r_mx}
\end{eqnarray}
  Observe that $l_{mx}$ is always less than $r_{mx} $ when $\pmax > p_{sw}$.

  In general, when  $l_{mx}  <  r_{mx} $, the optimizer along this boundary is obtained as in the proof of Theorem \ref{thm_Fco_positive} and equals (see Appendix for definitions)
\begin{eqnarray}
  U^*_{mx} &=& \hspace{-2mm}U_\sV (\pmax, q^*(\pmax) )\nonumber\\ 
  \mbox{ where } q^*(\pmax) 
  &=& \hspace{-2mm}\max  \{  l_{mx}  ,  \min\{  r_{mx}, h (\pmax) \} \}. \label{Eqn_opt_max_util}
\end{eqnarray}

We are left with two more sub-regimes,  one where the opponent's optimal price equals $\p_{mx}$ in \eqref{Eqn_opt_policy_Mj} and the other where opponent $M_j$ is made to operate at par. In the latter case, the optimal utility of $M_j$  exactly equals zero. We analyze both of them together in the following sub-section.

\subsection  {Opponent operates at par or the price saturates} 
From \eqref{Eqn_opt_policy_Mj} and \eqref{eqn_util_co-exist_given_pq},  when the optimal price of the opponent saturates at $\p_{mx}$, then   the utility of $\V_i$ coalition modifies to the following:
    \begin{eqnarray*}
      U_{st}(p,q) &:=& U_\sV(p,q)\\
      &= & \hspace{-2mm}\left(\dbar_\sMi(1+\varepsilon^2) + \varepsilon\dbar_\sMj - \alpha_\sMi p \right)\left(p- C_\sMi - C_\sS\right) \\
      &&+ \varepsilon\left(\alpha_\sMi p -\dbar_\sMi \right)\left(q- C_\sS \right) - O_\sMi - O_\sS.
    \end{eqnarray*}
    The set of $(p,q) \in {\cal F}_{co}$ where such a saturation occurs is given by (see \eqref{Eqn_feasible_Regioin_Mj} \eqref{Eqn_phi_st_p}):

    \vspace{-4mm}
    {\small\begin{eqnarray}
        {\cal F}^{st}_{co} &=& \left \{ (p,q) \in {\cal F}_{co} :  \p^* (p,q) =\frac{\dbar_\sMj + \varepsilon 
    \dbar_\sMi}{\alpha_\sMj}  \right \} \nonumber \\ &=&     \left \{ (p,q) \in {\cal F}_{co} :  q > \phi(p)  \mbox{ and }  q \le \theta(p) \right \} \label{Eqn_Fco_Saturate} .
    \end{eqnarray}}
Comparing section wise,  once again across $q$, one can easily verify that 
    $$
    U_{st}(p, q) \le U_{st} (p, \theta(p) ) \mbox{ for all } p \mbox{ such that  }  (p, \theta(p) ) \in {\cal F}_{co}^{st}. 
    $$
   Further,  it is not difficult to see that if there exists a $p$ such that $(p,q) \in {\cal F}_{co}^{st}$, then $(p,\theta(p)) \in {\cal F}_{co}^{st}$. 
  Thus the optimal co-existence utility in ${\cal F}_{co}^{st}$ is given by the optimal across all points in which the opponent operates at par, i.e., in ${\cal F}_{co}^{st} \cap \{ (p, \theta(p) ) \}.$  As a result, \textit{ the optimal across all the co-existence points when the opponent operates at par or when its price saturates can be derived in a combined manner ---  the relevant optimization problem is given by}:
  \begin{eqnarray*}   
 U_{pr}^* := \max_{ (p, q) \in {\cal F}_{co} : q = \theta(p) }  U_{\sV} (p,q).
  \end{eqnarray*}

Towards solving the above optimization problem, we need to proceed separately depending upon  the sign of  $ (p_{sw}-p)  $    (see \ref{Eqn_feasible_Regioin_Mj}). The following optimization problem is relevant for $p \le p_{sw}$ 
\[
\scalebox{0.75}{$
\begin{aligned}
    \max_{ p \le \min\{p_{sw}, \pmax\} }
    &\Bigg( \left(\dbar_\sMi + \varepsilon\dbar_\sMj - \alpha_\sMi(1-\varepsilon^2)p - \varepsilon\sqrt{\alpha_\sMj O_\sMj}\right)\left(p - C_\sMi - C_\sS\right) \\
    &\hspace{-15mm}+ 
    \frac{\sqrt{\alpha_\sMj O_\sMj}\left(\dbar_\sMj + \varepsilon\alpha_\sMi p - \alpha_\sMj C_\sMj - \alpha_\sMj C_\sS - 2\sqrt{\alpha_\sMj O_\sMj}\right)}{\alpha_\sMj}  - O_\sMi - O_\sS
    \Bigg).
\end{aligned}
$}
\]
The optimizer of the above by strict concavity is  at $ p^{1, *} $ given below:

\[
\scalebox{0.85}{$
\begin{aligned}
    \min \Bigg\{ p_{sw}, \pmax, & \frac{(C_\sMi + C_\sS)}{2} \\
    &+ \frac{\left( \dbar_\sMi + \varepsilon\dbar_\sMj - \varepsilon\sqrt{\alpha_\sMj O_\sMj} + \frac{\varepsilon\alpha_\sMi \sqrt{\alpha_\sMj O_\sMj}}{\alpha_\sMj} \right)}{2\alpha_\sMi(1-\varepsilon^2)} \Bigg\}.
\end{aligned}
$}
\]
And the second optimization for $p > p_{sw}$ is given by the following and is applicable only when $p_{sw} \le  \pmax$:
\scalebox{0.7}{$
\begin{aligned}
    \max_{ p_{sw} \le p \le \pmax }
    & \Bigg( - O_\sMi - O_\sS + \left( \dbar_\sMi(1+\varepsilon^2) + \varepsilon\dbar_\sMj - \alpha_\sMi p \right) \left(p - C_\sMi - C_\sS\right) \\
    & \quad \hspace{-20mm} + \left( \varepsilon\alpha_\sMi p - \varepsilon\dbar_\sMi \right) \left( \frac{\left( \dbar_\sMj + \varepsilon\dbar_\sMi - \alpha_\sMj C_\sMj - \alpha_\sMj C_\sS \right)\left( \varepsilon\alpha_\sMi p - \varepsilon\dbar_\sMi \right) - \alpha_\sMj O_\sMj}{\alpha_\sMj (\varepsilon\alpha_\sMi p - \varepsilon\dbar_\sMi)} \right)  
   \Bigg).
\end{aligned}
$}

    The optimizer of the above by strict concavity is at   $p^{2, *}$, given below: 
    \vspace{4mm}
\scalebox{0.59}{$
\begin{aligned}
    &\max \left\{ p_{sw}, \ \min \left\{ \pmax, \frac{\dbar_\sMi(1+\varepsilon^2) + \varepsilon\dbar_\sMj + \alpha_\sMi(C_\sMi + C_\sS) + \frac{\varepsilon\alpha_\sMi}{\alpha_\sMj}(\dbar_\sMj + \varepsilon\dbar_\sMi - \alpha_\sMj C_\sMj - \alpha_\sMj C_\sS)}{2\alpha_\sMi} \right\} \right\}.
\end{aligned}
$}

 \vspace{-2mm}  
In all, the optimal value in the combined sub-regime has optimal point where the out-house $M_j$ is forced to operate at par, and  is given by:

\vspace{-4mm}
{\small\begin{eqnarray} 
U_{pr}^* = \max\Bigg \{ U_\sV (p^{1,*}\theta(p^{1,*}), U_\sV (p^{2,*}, \theta(p^{2,*}) \indc{p_{sw} > \pmax }   \Bigg \}.\hspace{1mm}\label{Eqn_opt_util_par} 
\end{eqnarray}}

We finally have the following result using  Theorem \ref{thm_Fco_positive} and optimal utilities of  \eqref{Eqn_opt_loss_util}, \eqref{Eqn_opt_max_util}, \eqref{Eqn_opt_util_par}:
\begin{theorem}
\label{Thm_all_in_one}
    Assume {{\bf A.1-2}}.  We then have 
    \begin{eqnarray*}
        U^*_{co} 
      =  \max \Bigg \{  U(p_{co}^*, q_{co}^*) \indc{ (p_{co}^*, q_{co}^*) \in {\cal F}_{co}^+ }, \  \ U_{pr}^*,\\    U_{ls}^{*} \indc{\psi(0) < \pmax},   \ U^*_{mx} \indc{ l_{mx}  <  r_{mx} }   \Bigg \}. 
    \end{eqnarray*}
    
\end{theorem}
{\bf Remarks:} Thus the optimal operating point for $\Vi$ is: (a) either in the interior of ${\cal F}_{co}^+$, briefly referred to as 'operate both profitably' --- both the manufacturers derive non-zero profits at such points;
(b) or when the out-house is forced to operate at par, such a regime is briefly referred to as 'operate at par' --- here    $\V_i$ quotes an optimal point $(p^*,q^*)$ at boundary $p^*=\theta(q^*)$;
(c) or in    regime where its in-house unit incurs losses, briefly referred to as 'operate at losses' --- such a regime is non-empty only when $\psi(0) < \pmax$; 
(d) or the optimal price quoted by its in-house $M_i$ equals $\pmax$, referred to as 'operate at max' --- such a regime is non-empty only when $l_{mx} < r_{mx}$ (see \eqref{Eqn_r_mx}).

\vspace{2mm}
 \section{Comparison Analysis}
 We now compare the different regimes to identify the beneficial regimes for the given market conditions. 
 To begin, we consider the following term  (see \eqref{Eqn_psi_q})
\begin{eqnarray*}
    \psi(0) - p_{mx}
    &=& \frac{  \varepsilon^2  \dbar_\sMi - \varepsilon  (1-\varepsilon^2) \dbar_\sMj + \varepsilon \alpha_\sMj  C_\sMj   }{\alpha_\sMi (2-\varepsilon^2)}.
\end{eqnarray*}
Thus when $\varepsilon d_\sMi >    (1-\varepsilon^2) \dbar_\sMj -  \alpha_\sMj  C_\sMj$,    operating its in house production unit at losses is never a good option. 

Interestingly this does not depend either upon its reputation nor upon its production capacity. For further analysis we prove the following, whose proof is in Appendix. 
\begin{lemma}
\label{lem_comp}
 \textit{If $(8-6\varepsilon^2)\alpha_\sMi \alpha_\sMj - \varepsilon^2(\alpha_\sMi^2 + \alpha_\sMj^2) < 0,$ then the  $\Vi$ coalition finds it beneficial to either operate at loss or at par or at maximum price.}  \eop 
\end{lemma}

Thus under the above assumptions, it is not optimal for  $\Vi$ coalition to operate at a point where both the manufacturers derive non-zero profits, unless it is optimal to quote   maximum possible price $\pmax$ for  in-house products.

 Further,   for any given set of parameters excluding $\varepsilon$, 
there exists a $\bar \varepsilon  < 1$, such that for all $\varepsilon \ge \bar \varepsilon$ (while the other parameters  are kept fixed), it is not optimal to operate both profitably  ---  
  the term $(8-6\varepsilon^2)\alpha_\sMi \alpha_\sMj - \varepsilon^2(\alpha_\sMi^2 + \alpha_\sMj^2) $ of Lemma \ref{lem_comp}, converges to a negative value  as $\varepsilon \to 1$.  
 Using this, we finally derive the following result, whose proof is in Appendix A: 
 \begin{lemma}\label{lem_compr}
\textit{ (i) For any given set of parameters excluding $\varepsilon$, 
there exists a $\bar \varepsilon  < 1$, such that for all $\varepsilon \ge \bar \varepsilon$  it is either beneficial to operate at par or at maximum price. (ii)  Further if, }

\vspace{-3mm}
{\small
\begin{eqnarray}
(\alpha_\sMj - \alpha_\sMi)\dbar_\sMi + (2\alpha_\sMi + \alpha_\sMj)\dbar_\sMj +\alpha_\sMi \alpha_\sMj(C_\sMj - C_\sMi) \nonumber  
\\ 
&
\hspace{-83mm} < \ 
2\sqrt{2}\alpha_\sMi \sqrt{ (\dbar_\sMj)^2 - \alpha_\sMj O_\sMj  }\label{Eqn_cond_forat_max} \hspace{1mm}
\end{eqnarray}} 
\textit{it is beneficial to operate at max (for all such $\varepsilon$) with the optimal point being $(\pmax, h(\pmax) )$.}
\ignore{, or,
\vspace{3mm}
\scalebox{0.65}{$
\begin{aligned}
 \left( \frac{\dbar_\sMi + \varepsilon \dbar_\sMj}{\alpha_\sMi}, \ \frac{\varepsilon (\alpha_\sMj+\alpha_\sMi) \frac{\dbar_\sMi + \varepsilon \dbar_\sMj}{\alpha_\sMi} - \varepsilon\alpha_\sMj \left(C_\sMi + C_\sS\right) + \left(\dbar_\sMj - \alpha_\sMj C_\sMj + \alpha_\sMj C_\sS\right)}{2 \alpha_\sMj} \right) .
\end{aligned}
$}}
\textit{(iii) If  \eqref{Eqn_cond_forat_max}   is negated, then it is optimal to operate at par with $(\pmax, \theta(\pmax) )$.\eop}
\ignore{, or,
$$
 \left  (  \frac{\dbar_\sMi + \varepsilon \dbar_\sMj}{\alpha_\sMi}, \   \frac{\dbar_\sMj + \varepsilon\dbar_\sMi - \alpha_\sMj C_\sMj  }{\alpha_\sMj} -\frac{   O_\sMj}{ \varepsilon^2 \dbar_\sMj}  \right ).
$$.} 

\end{lemma}

Thus when the two units are identical, or even if the  in-house is inferior to an extent that still satisfies the condition of part (ii), $\Vi$ can compel the out-house unit to operate at par, once the substituitability factors are sufficiently high. 

A  result in a similar direction follows again by Lemma \ref{lem_comp}, even when the reputation factors   are  different.  There exists a threshold $\bar \gamma $ and  when, 
\begin{eqnarray*}
    \frac{\max \{ \alpha_\sMi, \alpha_\sMj \}  }{  \min \{ \alpha_\sMi, \alpha_\sMj \} }  > \bar \gamma, 
\end{eqnarray*}
operate both profitably is not an optimal choice. 
Interestingly   threshold $\bar \gamma$  depends  only upon $\varepsilon$ and not on other  parameters. We now consider some numerical example to derive further insights regarding  the optimal choice under such asymmetric conditions.

\vspace{2mm}
\section{Numerical Observations}
 By numerically computing the quantities of Theorem \ref{Thm_all_in_one},  we derive the required numerical inferences. 
 We set  $C_\sMi = C_\sMj = 4$, $C_\sS = 3$ and $O_\sMi= O_\sMj = O_\sS = 10$ and vary  other factors to   investigate the impact of different market conditions.

\begin{figure*}[h]
\vspace{ 15 mm}
\hspace{-9 mm}
     \centering
    \begin{minipage}{3.3cm}
     \vspace{-3mm}
 \includegraphics[trim = {0.5cm 1.5cm 0.5cm 4.5cm}, clip, scale = 0.18]{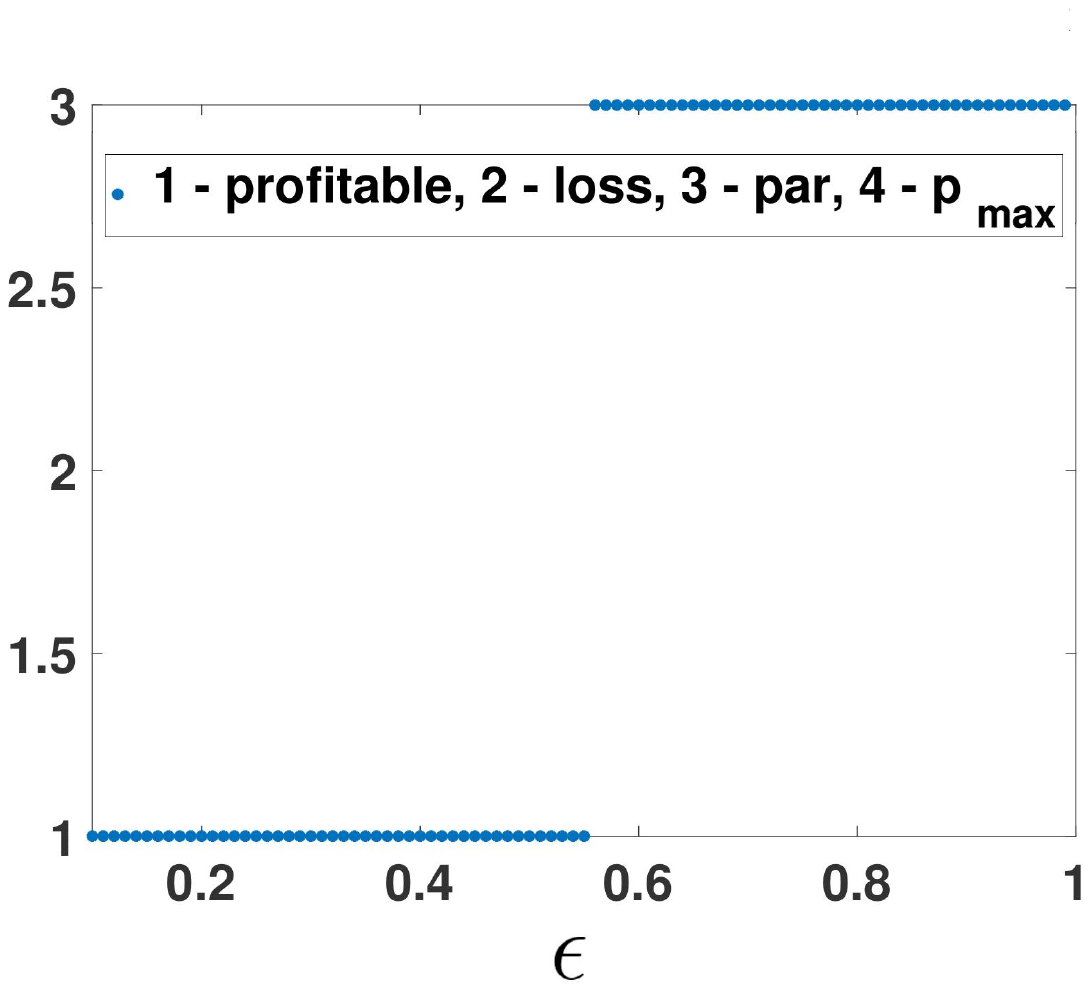}
\vspace{-14mm}
  \caption{\small{Symmetric case}} 
   \label{fig:sym_manu}
     \end{minipage}
     \hspace{1 cm}
         \begin{minipage}{3.4cm}
 \includegraphics[trim = {7cm 8cm 3.5cm 8cm}, clip, scale = 0.34]{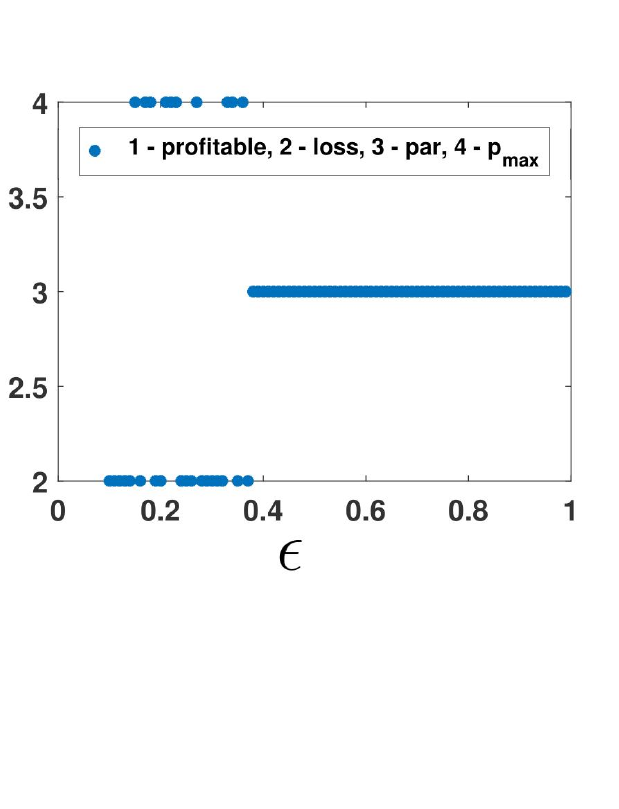}
 \vspace{-20mm}
    \caption{\small{Inferior in-house}}
   \label{fig:manuiinferior}
     \end{minipage} 
     \hspace{0.6cm}
         \begin{minipage}{3.5cm}
        \vspace{-6mm}  
\includegraphics[trim = {1cm 6.5cm 1cm 3.5cm}, clip, scale = 0.2]{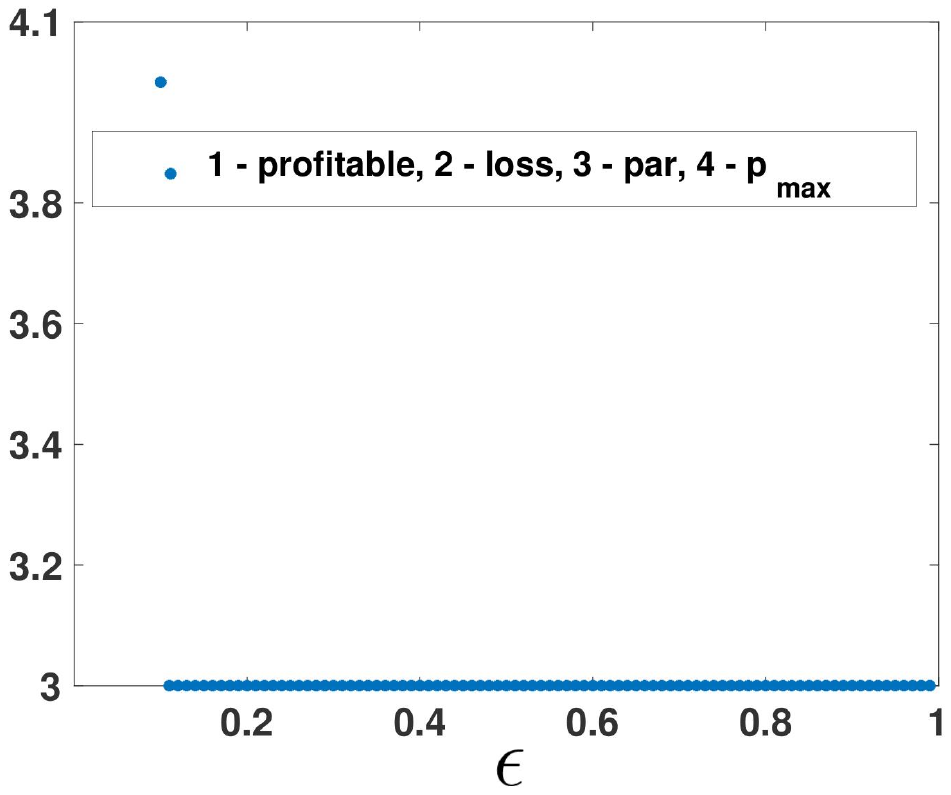}
 \vspace{-7mm}
   \caption{Non comparable.} 
    \label{fig:non-comparable}
     \end{minipage}
 \end{figure*}

In the first experiment provided in Figure \ref{fig:sym_manu}, we consider  a completely symmetric scenario --- 
  the manufacturers have equal market powers, basically   equal market potentials and price sensitivity parameters (we set $\dbar_\sMi=\dbar_\sMj = 100$ and $\alpha_\sMi = \alpha_\sMj = 0.1$). 
  We obtain the optimal configuration as a function of $\varepsilon$, the substitutability  factor.
  As seen from Figure \ref{fig:sym_manu}, for the symmetric case with smaller $\varepsilon$, the optimal choice for $\Vi$ is to operate both profitably (represented by $1$ on y-axis). On the other hand, when  $\varepsilon$ is high, the optimal configuration is to compel out-house $M_j$ to operate at par (represented by $3$ in figure).  
  

In Figure \ref{fig:manuiinferior},
a second experiment with inferior in-house manufacturer is considered. We set $\dbar_\sMi =10 \ll  \dbar_\sMj = 100$ and consider that  $\alpha_\sMi =0.1 \gg  \alpha_\sMj = 0.001$, basically the
  in-house manufacturer has smaller market potential as well as higher price sensitivity factor. 
For this case, it is again optimal to operate-at par for higher $\varepsilon$. However for small values of $\varepsilon,$  the optimal configuration  is either operate at loss or to operate at the maximum price.  


In Figure \ref{fig:non-comparable},We set $\dbar_\sMi =10 \ll  \dbar_\sMj = 100$ and consider that  $\alpha_\sMi =0.001 \gg  \alpha_\sMj = 0.1$, basically we consider   non comparable manufacturers. Here   the  market potential of the in-house manufacturer is smaller,   while the price sensitivity of the out-house manufacturer is larger.  
Interestingly for this case, operate at par is optimal for almost all the values of $\varepsilon$.

In all, irrespective of the market capacities of both the units, we found that the supplier has managed to compel the out-house to operate at par, once $\varepsilon$ is sufficiently high (like at least 0.5). 




\vspace{2mm}
 \section{Conclusions}  
 We have investigated the optimal choices of a supplier supplying material to a manufacturer,  which additionally has an in-house production unit. We considered a market model with dedicated customer bases, which however   are influenced by the prices of  both the  production units.  The presence of in-house unit facilitates the supplier to gain control over the entire supply chain. For instance, at an extreme, even with an inferior in-house production unit  (i.e., with smaller market power), it could manage to compel the   
 out-house (with a much larger customer base) to operate at par, there by avoiding  the downstream monopoly.

 This initial study inspires many more open questions for future investigation.  Comparison of the supplier gains with and without in-house unit?  What happens if a more realistic dynamic model including inventory control and fluctuating demands is considered?

 \ignore{

 {\color{red} We have explored the optimal choices for a supplier providing materials to a manufacturer that also operates an in-house production unit. Our analysis considers a market model with distinct customer bases, where demand for each unit is influenced by the pricing of both. The in-house unit enables the supplier to exert control over the entire supply chain. For example, even if the in-house unit has relatively limited market power, it can still influence the larger out-house operation to perform competitively, thereby preventing downstream monopolistic control.

This initial study opens the door to several potential research avenues. For instance, how does the supplier’s profit compare with and without an in-house unit? How would a more dynamic model, incorporating inventory control and fluctuating demand, impact the results? These questions present intriguing directions for the future study.}}
 
 \ignore{ 
 which are sensitive to the quoted prices and a fraction of unsatisfied

 as well as outsources raw material to a downstream manufacturer under different market regimes. We observe that when the supplier's in-house manufacturer and the out-house manufacturer are identical or nearly identical, it is beneficial for the supplier to both supply raw material to out-house manufacturer and compete with it in downstream market gaining profits from both the operations. This happens when the manufacturers are not substitutable. On the other hand in this case when these manufacturers are substitutable, then the supplier finds it beneficial to operate the out-house manufacturer at par.   

\ignore{
 {\color{blue}
(ii)  When both manufacturer are identical or are nearly identical, we have either co-existence (when $\varepsilon$ is low). or at par (when $\varepsilon$ is high).

(iii) Co-existence not possible at $\varepsilon$ close to 1 (?)

(iv) When $\alpha_\sMj \to 0$ it is optimal to  operate at loss.

(v) When $\alpha_\sMi \to 0$,we have } 
}
}

 \bibliographystyle{ieeetr}

\begin{thebibliography}{00}
\bibitem{wang2013advantage}Wang, Y., Niu, B. and Guo, P., 2013. On the advantage of quantity leadership when outsourcing production to a competitive contract manufacturer. Production and Operations Management, 22(1), pp.104-119.\\
\bibitem{arya2007bright}Arya, A., Mittendorf, B. and Sappington, D.E., 2007. The bright side of supplier encroachment. Marketing Science, 26(5), pp.651-659.\\
\bibitem{nystedt2007acer}Nystedt, D., 2007. Acer passes Lenovo in Q1, next up.\\
\bibitem{ha2022supplier}Supplier encroachment, information sharing, and channel structure in online retail platforms\\
\bibitem{yoon2016supplier}Yoon, D.H., 2016. Supplier encroachment and investment spillovers. Production and Operations Management, 25(11), pp.1839-1854.\\
\bibitem{wadhwapartition}Wadhwa, G., Walunj, T.S. and Kavitha, V., Partition-Form Cooperative Games in Two-Echelon Supply Chains.\\
\bibitem{williamson1971vertical}Williamson, O.E., 1971. The vertical integration of production: market failure considerations. The American economic review, 61(2), pp.112-123.\\
\bibitem{zheng2021willingness}Zheng, X.X., Li, D.F., Liu, Z., Jia, F. and Lev, B., 2021. Willingness-to-cede behaviour in sustainable supply chain coordination. International Journal of Production Economics, 240, p.108207.\\
\bibitem{amirnequiee2024navigating}Amirnequiee, S., Pun, H. and Naoum-Sawaya, J., 2024. Navigating supplier encroachment: Game-theoretic insights for outsourcing strategies. European Journal of Operational Research, 319(2), pp.557-572.\\
\bibitem{simchi1999designing}Simchi-Levi, D., Kaminsky, P. and Simchi-Levi, E., 1999. Designing and managing the supply chain: Concepts, strategies, and cases. New York: McGraw-hill.\\
\bibitem{TR}{http://www.ieor.iitb.ac.in/files/faculty/kavitha/SC.pdf}
 \end{thebibliography}

\vspace{-4mm}

%
{
\section{Appendix A}

\noindent {\bf Proof of Theorem \ref{thm_Fco_positive}:}  
From  \eqref{Eqn_phi_st_p}, 
 $\phi(p) < 0$, when $\alpha_\sMi p > \dbar_\sMj + 2 \varepsilon \dbar_\sMi - \alpha_\sMj C_\sMj$. For such $p$,    $(p,q) \notin {\cal F}_{co}^+$ for any   $q$ (see \eqref{Eqn_Fco_plus}). 
 Also, from \eqref{Eqn_psi_q}, $p \le \psi(q)$ if and only if $q \ge \psi^{-1} (p)$.
 Thus a more direct  representation of ${\cal F}_{co}^+$   \eqref{Eqn_Fco_plus} is given by:

\vspace{-4mm}
{\scriptsize
\begin{eqnarray}\label{eqn_cal_F_co_+}
    {\cal F}^+_{co} &=& \hspace{-2mm} \left \{ (p, q)    :  0 \le   p \le \bar{p} (q)  \mbox{ and } \max\left \{0, \psi^{-1}(p) \right \} \le  q \le {\bar q}(p)  \right \},     \hspace{4mm} \\
    {\bar p} (q) &:=&
    \min \left \{\pmax, \psi(q), \left (\phi\right )^{-1}(0)  \right  \}  \nonumber \\
    &= & \min \left \{ \frac{\dbar_\sMi + \varepsilon \dbar_\sMj}{\alpha_i}, \psi(q), \frac{\dbar_\sMj + 2\varepsilon\dbar_\sMi -\alpha_\sMj C_\sMj}{\varepsilon\alpha_\sMi}   \right  \}. \nonumber  \mbox{ and }  \\
    {\bar q}(p) &:=& \min \left \{\theta (p), \phi (p) \right  \} \stackrel{a}{=} \left \{ 
    \begin{array}{lll}
       \theta (p)  &  \mbox{ if } p  \le \frac{\dbar_\sMi}{\alpha_\sMi} + \frac{\sqrt{\alpha_\sMj O_\sMj}}{\varepsilon\alpha_\sMi} \\
      \phi (p)    & \mbox{ else. }
    \end{array} 
    \right . 
\end{eqnarray}}
(by direct computations using \eqref{Eqn_phi_st_p} and \eqref{Eqn_opt_policy_Mj} one can verify equality `$a$'). 

The function
$U_\sV$ is continuous and ${\cal F}_{co}^+$ is bounded (as $\pmax < \infty$), thus we have an optimizer for \eqref{eqn_vc_opt_f_co_+}.

  Define p-sections $\S_p := {\cal F}_{co}^+ \cap \{(p, q): q \ge 0\}$ lines for each $p \le \pmax$. 
The idea is to find sub-optimizers in each 
  $\S_p$ and then find the global optimizer. 
  Towards this goal, first note that
     the function $U_\sV$ in  ${\cal F}_{co}^+$ matches with  the   `unconstrained' function $U$ given in equation \eqref{eqn_util_co-exist_uc}, which can be rewritten as (see \eqref{Eqn_ws} for definitions):

\vspace{-3mm}
{\scriptsize
\begin{eqnarray}\label{eqn_Util_w}
    U(p,q) \ = \ w_1 p^2 + w_2 pq + w_3 q^2 + w_4 p + w_5 q + w_6, \mbox{ with }    \\
& \hspace{-76mm} w_6 \ = -\left(\dbar_\sMi + \frac{\varepsilon\left(\dbar_\sMj + \alpha_\sMj C_\sMj \right)}{2}\right)\left(C_\sMi + C_\sS \right) - \left(\frac{\dbar_\sMj - \alpha_\sMj C_\sMj}{2}\right)C_\sS.
    \nonumber 
\end{eqnarray}}    
The   second derivative $\nicefrac{\partial^2 U}{\partial^2 q} = w_3 < 0$ for all $(p,q)$. 
 %
     Thus for any $p$ with $\S_p \ne \emptyset$,  the     sub-optimizer of  the sub-optimization problem $ \max_{q : (p,q)\in \S_p} U(p,q)$ is  unique by strict concavity  and equals,
  
  \vspace{-4mm}{\small\begin{eqnarray}\label{eqn_q_star_p}
      q^*(p) &:=& 
     \max\{l(p),  \min \{ h(p),  {\bar q}(p)  \},  \mbox{ where }\nonumber \\
     h(p) &:=& - \frac{w_2 p + w_5} {2 w_3}  
  \end{eqnarray}}%
     is the  `unconstrained'  optimizer  of  $U (p, \cdot)$ over $\{  q  \in {\cal R} \}$, 
      ${\bar q}(p)$ is the right  boundary point  and   $l(p) := \{0, \psi^{-1}(p) \} $ is the left boundary point  of $\S_p$ (see \eqref{eqn_cal_F_co_+}).   

 Define  

\vspace{-4mm}
{\tiny\begin{eqnarray}
         \bar p &:=& \bar p(0) = \min \left \{\psi(0), \pmax, \ \phi^{-1} (0) \right  \}  \label{Eqn_pbar}\\   \nonumber 
         &
      & \hspace{-10mm}= 
     \min \left \{ \frac{   2 \dbar_\sMi + \varepsilon  \dbar_\sMj + \varepsilon \alpha_\sMj  C_\sMj }{\alpha_\sMi (2-\varepsilon^2)},  \frac{\dbar_\sMi+\varepsilon \dbar_\sMj }{\alpha_\sMi}, \frac{\dbar_\sMj + 2\varepsilon\dbar_\sMi -\alpha_\sMj C_\sMj}{\varepsilon\alpha_\sMi} \right  \}.   
 \end{eqnarray}}

 From \eqref{Eqn_phi_st_p}, \eqref{Eqn_psi_q} and \eqref{eqn_cal_F_co_+} and  with   $p \le  {\bar p} (0)$,  we have  $\phi(p) \ge   0$   and so $   {\bar q}(p) \ge  0$ (as from \eqref{Eqn_feasible_Regioin_Mj},  $\theta(p') >0$ for any $p'$) and $\psi^{-1}(p) \le 0$; hence  for all such $p$, we have $\S_p \ne \emptyset$ with left boundary $l(p) = 0$. Further   $h(p) >0$ for all $p$ by {\bf A}.2 and thus \eqref{eqn_q_star_p}  equals:

\vspace{-3mm}
{\tiny\begin{eqnarray}
    q^*(p) = \left \{
    \begin{array}{ll}
  \min \{ h(p),  {\bar q}(p) \}       &  \mbox{ if } p \le \bar p \\ \\
\max \{ l(p),   \min \{ h(p),  {\bar q}(p) \}   \}         & \mbox{ else, i.e., if and only if,  } {\bar p}  < p  < \pmax   
    \end{array}
    \right . 
    \label{Eqn_qstar}
\end{eqnarray}}
 
We have `if and only if' in the last line of \eqref{Eqn_qstar}, because
when  $\bar p (0) < \psi(0)$: (i) either   ${\bar p} = \pmax$ and then clearly $\S_p = \emptyset$ for all $p > \bar p$;  (ii)   or 
 $\phi(\bar p) = 0$ and so $\phi(p) < 0$ (and so $\bar q(p) < 0$) for all $p > \bar p$,  and then again 
 $\S_p = \emptyset$ for all $p$.
 When $\bar p = \psi(0)$, for all $p > \bar p$ we have $l(p) = \psi^{-1}(p) > 0$. In all, we have $q^*  (p) > 0 $ for all $p$ with $\S_p \ne \emptyset$. 
 
From \eqref{Eqn_pbar} and {\bf A}.1  we have $\bar p > 0$, thus  
 there exists at  least one $p$ such that $\S_p \ne \emptyset$,
  and hence:
$$
\max_{(p, q) \in {\cal F}_{co}^+} U_\sV(p,q) = \max_{p \le \pmax, \S_p \ne \emptyset}  U_\sV(p, q^*(p) ). 
$$
In other words,   the global optimizer of $U_\sV$ in ${\cal F}_{co}^+$ is among, 

\vspace{-4mm}
    {\scriptsize
    \begin{eqnarray} 
\nonumber
    \mathbb {L}^* &:= &\{ (p, q) : 0\le p \le \pmax, \S_p \ne \emptyset, q = q^*(p) \} \\
     &=& \bigg  \{ (p, q) :  0 \le  p \le   {\bar p}, \ \  \ \ \ \ \ \ \ q= q^*(p) = \min \{ h(p),  {\bar q}(p) \} \bigg  \}   \nonumber \\
&& \cup \bigg \{ (p, q) : {\bar p}  < p \le \pmax, \ \S_p \ne \emptyset, \ \ \  q = q^*(p) \bigg  \} . \label{Eqn_L_star} 
 \end{eqnarray}}
 Also since $q^*(p) > 0$ for all $\S_p \ne \emptyset$, we have ${\mathbb L}^* = {\mathbb L}^* \cap \{ (p, q) : q > 0\}. $
 Further proof is obtained   by proving that the mapping $\omega(p) := U(p, h(p))$ is either  concave or convex,  and this is continued in \cite{TR}.\eop 

 The remaining proofs of  Appendix A are also  in \cite{TR}.

 \end{document}